\documentclass[aps,amsart
amsproc,article,elsarticle,emulateapj]{revtex4}
\usepackage{graphicx}
\usepackage{amsmath}
\usepackage{csquotes}
\usepackage{xcolor}

\usepackage{siunitx}

\begin{document}


\title{  Quantum gravity phenomenology in quadrupolar Bose-Einstein condensates:  Finite-temperature thermodynamics and projected  bounds}

\author{  Asma Tahar Taiba$^{1}$ and Abdel\^{a}ali Boudjem\^{a}a$^{2}$}

\affiliation{$^1$  LPTHIRM, Department of Physics, Faculty of Sciences, University of Blida 1,  P.O. Box. 270, 09000, Blida, Algeria \\
$^2$ Department of Physics, Faculty of Exact Sciences and Informatics, 
and Laboratory of Mechanics and Energy, Hassiba Benbouali University of Chlef, P.O. Box 78, 02000, Chlef, Algeria.} 

\email {a.boudjemaa@univ-chlef.dz}

\date{\today}

\begin{abstract}

We study the thermodynamics of quadrupolar Bose-Einstein condensates in the realm of the generalized uncertainty principle (GUP) for the quest of understanding and testing quantum gravity (QG). Using both quadratic GUP (QGUP) and linear-quadratic GUP (LQGUP) which incorporate modified density of states, we calculate corrections to the condensed fraction, critical temperature, equation of state and other thermodynamic quantities. We find that the intriguing interplay of quadrupole-quadrupole interactions (QQIs) and the GUP may strongly modify the finite-temperature properties of the condensate, with the LQGUP generally predicting larger corrections than the QGUP, although their relative magnitudes depend on the
chosen parameters and the validity of the perturbative expansion.
We further estimate the sensitivity of  the critical temperature to GUP-induced effects by imposing an assumed fractional temperature resolution which provides projected upper bounds on the GUP deformation parameters.
 Our findings demonstrate that quadrupole-quadrupole interacting Bose gases provide a promising platform for testing GUP-induced deviations from standard quantum mechanics and for probing QG phenomenology through measurements 
 of finite-temperature many-body observables.   They also  motivate future studies using realistic system-specific parameters.

\end{abstract}

	\maketitle

\section{Introduction}

The GUP  emerges from many approaches to QG, predicts the existence of a minimum length and maximum momentum, impacting the structures of phase space through the deformation of the density of momentum states (see e.g. \cite{Taw,Boss}).
Over the past three decades, there has been remarkable progress in studying GUP-induced corrections in the thermodynamics of ideal Bose and Fermi gases 
(see e.g., \cite{Fit, VaK,Zhang, Cast, Li, Sanj, Das} and references therein),  black-hole  physics (see e.g.  \cite{Chen, Chen1,Feng, Ong, Vag, Kouwn,Barm, Ior, Blach, Kanz, Zhou}),  
and weakly interacting Bose-Einstein condensates (BECs) with  contact and dipolar interactions \cite{Boudj,Boudj2}.  
In such systems the GUP manifests itself  through corrections to the equation of state (EoS), critical temperature, condensate fraction, and superfluid fraction.
Most recently, we have derived a new GUP formulation based on the generalized Gross-Pitaevskii equation describing the behavior of quantum self-bound droplets \cite{Boudj1}. 

Although these analyses have established a useful framework for incorporating QG effects into many-body physics, they have practically focused on noninteracting systems or atoms with
isotropic short-range interactions, they do not address an important question: how do QG-induced phase deformations interact with long-range and many-body quantum fluctuations?
Since the thermodynamic observables depend on both quasiparticles spectrum and the density of available momentum states, the response to the GUP is expected to depend on
the nature of the underlying interactions.

Quadrupolar BECs which have become an active area of research \cite{Bhongale2013,Lahrz2014,Li2013,Pikovski2014,Lahrz2015}, provide an ideal platform for addressing this issue and extending the above studies. 
They constitute a novel class of quantum fluids in which particles interact via anisotropic and long-range QQIs. 
Such an additional degree of anisotropy gives rise to fascinating many-body phenomena that are absent in ideal Bose gases 
whose thermodynamics is determined solely by quantum statistics, and the conventional weakly-interacting BECs, where the interactions are isotropic and short-range.
Quadrupolar molecules such as Cs$_2$ \cite{Web} and Sr$_2$~\cite{Stel,Rei}, together with the proposal of species carrying both electric and magnetic dipole moments~\cite{Lepers}, make the anisotropic QQI experimentally accessible and broadly tunable.
Very recently, one of us has developed a Bogoliubov theory describing nonequilibrium quench of disordered quadrupolar BECs \cite{Boudjemaa2025}.

Compared with dipolar condensates \cite{Pfau, Carr,Baranov}, the complex angular dependence and a longer range character may considerably alter the Bogoliubov excitation spectrum, 
collective mode, and the stability conditions, hence provide distinct finite-temperature thermostatistic properties.
In the context of the GUP, quadrupolar BECs are particularly appealing because thermodynamic quantities are determined by momentum-space integrals over the quasiparticle spectrum.
Since QQIs significantly reshape such a spectrum, they may modify the sensitivity of observables to GUP-induced deformation of phase space. 
Therefore, these exquisite characteristics of quadrupolar structures make them an informative testbed for investigating QG corrections 
than ideal gases and weakly-interacting condensates with a pure contact interaction or even dipolar BECs.





In this paper, we study the finite-temperature thermodynamics of homogeneous quadrupolar BECs within the framework of the GUP.
We adopt the effective phase-space formulation in which the microscopic interaction Hamiltonian and the Bogoliubov excitation spectrum remain unchanged, 
while the quantum-gravity correction is incorporated through the GUP-modified density of states. Accordingly, the standard momentum-space integration measure is replaced by
\begin{equation} \label{GUPFor}
\frac{d^{3}p}{(2\pi \hbar)^3}
\longrightarrow \frac{d^{3}p}{(2\pi \hbar)^3\mathcal{J}(p)},
\end{equation}
where the Jacobian $\mathcal{J}(p)$ relies on the chosen GUP model. Here we apply two popular models  namely: quadratic GUP (QGUP)  which was first proposed in Ref.~\cite{Kempf} 
and linear-quadratic GUP (LQGUP)  proposal  which was first  discussed in \cite{Das}.
The formulation (\ref{GUPFor}) preserves the established many-body description of the condensate while consistently accounting for QG effects through the deformation of phase space.

Starting from the  standard Bogoliubov spectrum combined with GUP-modified phase-space measure, we derive analytical formulas for thermal depletion, critical temperature and the condensed fractions. 
The grand potential and the EoS are obtained through the grand partition function with quasiparticle energy.  
Other thermodynamic quantities  such as  the internal energy, heat capacity, and entropy  are also computed by applying standard thermodynamic relations.
The obtained results demonstrate that the finite-temperature behavior of quadrupolar BEC differs substantially from that of an ideal Bose gas  \cite{VaK,Zhang, Cast, Li,Das} and  weakly interacting condensates with contact interactions 
\cite{Boudj}, while the magnitude and sign of the additional GUP corrections depends on the adopted deformation model and parameter regime. 
The findings reveal also  that  the LQGUP predicts a stronger modification of the thermodynamics than the QGUP most likely due to extra linear momentum corrections, which modifies the phase-space measure more strongly.

To probe possible low-energy  manifestations of QG, it is useful to examine how GUP corrections affect experimentally accessible observables and assess the sensitivity of suitable physical systems to these modifications.
We then estimate the projected sensitivity to the GUP deformation parameters by assuming a specified  fractional resolution in the critical temperature measurement.  
These estimates are conditional on the adopted effective phase-space prescription, the system parameters, and  the validity of the perturbative expansion.
In the LQGUP framework, the linear and quadratic contributions may partially cancel, so constraints on the two deformation parameters should ultimately be assessed jointly. Our analysis thus provides a theoretical basis for investigating  the potential of 
ultracold gases with QQIs as probes of GUP-inspired modifications of quantum physics  \cite{VaK,Zhang,Cast,Li,Das,Boudj,Boudj2,Boudj1}.

\section{Theory of quadrupolar BECs} \label{BHMT}

We consider a homogeneous weakly interacting quadrupolar BEC  having a quadrupole moment $q$ separated by the distance vector $r$ and aligned via an external magnetic field $B$ 
which is assumed to be fixed along the $z$-axis.
The Hamiltonian of the system can be written in terms of the creation and annihilation operators $\hat a^\dagger_{\bf p}$ and $\hat a_{\bf p}$ as \cite{Boudjemaa2025}:
\begin{align}\label{ham1}
\hat H\!\!=\!\!\sum_{\bf p}\! E_p\hat a^\dagger_{\bf p}\hat a_{\bf p}\! 
+\!\frac{1}{2 {\cal V}}\!\!\sum_{\bf p,\bf k,\bf q}\!\, V ({\bf q})\, \!\hat a^\dagger_{\bf p\!+\!\bf q} \hat a^\dagger_{\bf k\!-\!\bf q}\hat a_{\bf k}\hat a_{\bf p},
\end{align}
where $E_p=p^2/2m$, ${\cal V}$ is a quantization volume,  and $V(\mathbf p)$ is the interaction potential in Fourier space \cite{Lahrz2015}: 
\begin{equation}\label{intP}
V(\mathbf p)= g +g_q \frac{4 \pi}{105 \,\hbar^2}  Y_4^0 (\theta)  p^2,
\end{equation}
where $\theta$ is the azimuthal angle between the vector ${\bf p}$ and the polarization direction, $g=4\pi \hbar^2 a/m$
corresponds to the short-range part of the interaction with $a$ being the scattering length,
$g_q=q^2/ (\sqrt{\pi} \epsilon_0)$ characterizes the strength of the QQI, $\theta$ is the polar angle, and $Y_ 4^0 (\theta)$ is the spherical harmonic.
Unlike contact interaction,  QQI of Eq.~(\ref{intP})  depends in direction of $\mathbf k$ and generally of its magnitude.
It changes its character from repulsive for $\theta \in \left[ 0, \theta_1 =\arccos ( \sqrt{15+2 \sqrt{30})/35} \right]$ and $\theta \in \left[\theta_2 =\arccos\sqrt{ (15-2 \sqrt{30})/35}, \pi/2 \right]$ 
to attractive for $\theta \in [\theta_1 ,\theta_2]$ \cite{Lahrz2015}.

Performing a canonical Bogoliubov transformation: $\hat a_{p}= u_{p} \hat b_{p}- v_{p} \hat b_{-p}^\dagger$, where  the quasiparticle annihilation, $\hat b_{\mathbf{p}} $, and creation, $\hat b_{\mathbf{p}}^{\dagger} $, 
operators obey the usual Bose commutation relations: $\left[\hat b_{\mathbf{p}}, \hat b_{\mathbf{p'}}^{\dagger}\right]=\delta_{\mathbf{p,p'}}$,
$\left[\hat b_{\mathbf{p}}, \hat b_{\mathbf{p'}}\right]= \left[\hat b_{ \mathbf{p}}^{\dagger}, \hat b_{\mathbf{p'}}^{\dagger}\right]=0$.
The Bogoliubov quasiparticle amplitudes are defined as:
\begin{equation}\label{Eq12}
 u_{ p}=\frac{1}{2}\left(\sqrt{\frac{\varepsilon_{ p}}{E_p}}+\sqrt{\frac{E_p}{\varepsilon_{ p}}}\right),\;\;\; v_{p}=u_{ p}-\sqrt{\frac{E_p}{\varepsilon_{ p}}},
\end{equation}
which are chosen to be real without loss of generality.
Within this, the bilinear form of the Hamiltonian:
$$\hat  H\!\!= E+ \sum\limits_{{\bf p}\neq0} \varepsilon_{p} \hat b^\dagger_{\bf p} \hat b_{\bf p},$$ 
where 
\begin{equation}\label{Bogspec}
\varepsilon_{p}= \sqrt{E_p^2 + 2ng E_p \left[1+ \frac{4 \pi \epsilon_{\text{qq}}}{105} Y_4^0 (\theta) \left(\frac{p \xi}{\hbar}\right)^2 \right]},
\end{equation}
is the Bogoliubov spectrum,  $\epsilon_{\text{qq}}= g_q/(g \xi^2)$, and  $\xi=\hbar/\sqrt{mng}$ is the healing length.
Note that    $\epsilon_{\rm qq} $ can be experimentally adjusted by tuning the scattering length $a$ near a Feshbach resonance, enhancing the anisotropic QQI relative to the contact interaction. 

For small momenta, $\varepsilon_{p}= c_s p$, with $c_s=\sqrt{gn/m} $ being the sound velocity, and $n=N/ {\cal V}$ is the density of the gas. 
Remarkably, the sound velocity is isotropic (independent on the direction of the vector $\mathbf p$) in stark contrast to the standard dipolar BEC.
However, in the high momenta limit ($p\rightarrow \infty $), the spectrum turns out to be given as: 
\begin{equation}\label{BogspecHT} 
\varepsilon_{p}= A (\theta) E_p,
\end{equation}
where $A (\theta)=\sqrt{1 + (16 \pi \epsilon_{\text{qq}}/ 105) Y_4^0 (\theta)}$  contains the angular dependence associated with the quadrupolar interaction. 
For a  weak-quadrupolar-coupling expansion $ \epsilon_{\text{qq}} \ll1$, one has   $A (\theta) \simeq 1 + (8\pi \epsilon_{\text{qq}}/ 105) Y_4^0 (\theta)$.
The stability and reality of high-momentum spectrum requires  the criterion:
$$
1 + (16 \pi \epsilon_{\text{qq}}/ 105) Y_4^0 (\theta)>0 , \qquad \forall \theta. 
$$
Interestingly, for quadrupolar BECs, the excitation spectrum at high momentum differs from that of an ideal Bose gas and weakly interacting dipolar and nondipolar BECs, it remains anisotropic and retains an interaction-induced self-energy even though the dominant term is the free-particle kinetic energy. 
Therefore, all high-momentum thermodynamics quantities such the free energy, EoS, thermal depletion may acquire this anisotropy shift rather than the isotropic standard BEC.
The physical origin of such a discrepancy is indeed more pronounced than that of dipolar BECs since QQI has different momentum dependence.

{\it We will set Boltzmann constant $k_B=1$ throughout the manuscript.}

The use of Eq.~(\ref{BogspecHT} )  in the thermodynamic integrals corresponds to the high-momentum regime of the Bogoliubov spectrum, for which 
\[
\frac{p^2}{2m}\gg gn,
\qquad\text{or equivalently}\qquad
p\xi\gg 1.
\]
For thermally populated quasiparticles, the characteristic momentum is determined by $p_T^2/ (2m) \sim T$, so that the relevant condition can be expressed as
\[
gn \ll T<  T_c.
\]
Near $T_c$, a self-consistent Hartree-Fock-Bogoliubov or renormalization group approach would be required to capture critical scaling.

The thermal depletion, $\tilde{n}_T$, can be obtained via $\tilde{n}={\cal V}^{-1}\sum_{\bf p} \langle\hat a^\dagger_{\bf p}\hat a_{\bf p}\rangle$. 
Using the Bogoliubov transformation, and the definition: $\langle \hat b^\dagger_{\bf p}\hat b_{\bf p}\rangle=N_{\bf p}$,
where $N_{\bf p}=[\exp(\varepsilon_p/T)-1]^{-1}$ are occupation numbers for the excitations, 
and $\langle \hat b^\dagger_{\bf p}\hat b^\dagger_{\bf p}\rangle=\langle \hat b_{\bf p}\hat b_{\bf p}\rangle=0$, we get
\begin{align}\label {norT}
\tilde{n}_T= \int \frac{d^3p} {(2\pi \hbar)^3} \left(u_p^2+v_p^2\right)N_p  
=\frac{1} { 4\pi^2 \hbar^3} \int_0^\pi \sin \theta d\theta \int_0^{\infty} \left(\frac{\varepsilon_{p}}{E_p}+\frac{E_p}{\varepsilon_{p}} \right) \frac{p^2 dp }{\exp(\varepsilon_p/T)-1}. 
\end{align}
At higher temperatures, the main contribution to this integral comes from single-particle excitations which have energies 
$\varepsilon_{p}= A (\theta) E_p$. Then $\left(u_p^2+v_p^2\right) = [ A (\theta)+A^{-1} (\theta)]/2$, and from Eq.~(\ref{norT}) with 
the standard Bose integral $\int_0^{\infty} x^{\,s-1}/ (e^{x}-1)\,dx\;=\;\Gamma(s)\,\zeta(s)$, where $\zeta(s)$ is the Riemann Zeta function and $\Gamma(s)$ is the gamma function, we obtain:
\begin{equation} \label{nth}
\tilde{n}^{(0)}_T = \frac{1}{4}\left(\frac{m T}{2 \pi \hbar^2}\right)^{3/2} \zeta(3/2) f_3(\epsilon_{\text{qq}}), 
\end{equation}
where  $f_3(\epsilon_{\text{qq}})=\int_{0}^{\pi} \left[A^{-1}(\theta )+A(\theta )\right] A^{-3/2} (\theta) \sin\theta d \theta$, 
which is special case $j=3$ of $f_j(\epsilon_{\text{qq}})=\int_{0}^{\pi} \left[A^{-1}(\theta )+A(\theta )\right] A^{-j/2} (\theta) \sin\theta d \theta$ (see Appendix A for more details). 
For $\epsilon_{\text{qq}} \ll 1$, $f_3(\epsilon_{\text{qq}}) \simeq 4$, then we reproduce the standard results of an ideal BEC 
$\tilde{n}^{(0)}_T = (m T/2 \pi \hbar^2 )^{3/2} \zeta(3/2)$. 

The thermal grand potential  is given in terms of the grand partition function, ${\cal Z}$, as:
 \begin{align}
\Omega_T =- T \ln {\cal Z} 
=\frac{T} { 4\pi^2 \hbar^3}  \int_0^\pi \sin \theta d\theta  \int_0^{\infty} p^2dp \ln \left(1- e^{-\varepsilon_{p}/T}\right). 
\end{align}
Thus, the thermodynamic quantities derived below describe the thermal quasiparticles sector and should not be interpreted as the complete EoS of the interacting Bose gas. 
However, the total thermodynamic potential contains, in addition the condensate mean-field  and zero-point quantum fluctuation contributions,
$\Omega=\Omega_{\rm MF}+\Omega_{\rm T}+\Omega_{\rm Q}$.

A straightforward calculation using  Eq.~(\ref{BogspecHT}), and $\int_0^{\infty} x^{\,s-1}\Big[-\ln\!\big(1-e^{-x}\big)\Big]\,dx\;=\;\Gamma(s)\,\zeta(s+1)$, gives
 \begin{equation} \label{Fth}
\Omega_T^{(0)} =- \frac{1}{2} T\left(\frac{m T}{2 \pi \hbar^2} \right)^{3/2} \zeta(5/2) h_{3}(\epsilon_{\text{qq}}),
\end{equation}
where $h_{3}(\epsilon_{\text{qq}})= \int_{0}^{\pi} A^{-3/2} \sin\theta d \theta$ which is special case $j=3$ of $h_j(\epsilon_{\text{qq}})=\int_{0}^{\pi}A^{-j/2} (\theta) \sin\theta d \theta$ 
(see Appendix A for more details). 
Clearly, in the absence of QQI, $\Omega_T^{(0)} $ reduces to its standard form.

The  thermal quasiparticle contribution to the  EoS can be derived via $P_T=-\Omega_T/V$ . Using  Eq.~(\ref{Fth}), we  immediately  get
\begin{equation}
P_T V=2  \frac{\zeta(5/2) h_{3}(\epsilon_{\text{qq}})}{\zeta(3/2) f_3(\epsilon_{\text{qq}})} \tilde{n}^{(0)}_T T \approx \frac{h_{3}(\epsilon_{\text{qq}})}{f_3(\epsilon_{\text{qq}})} \tilde{n}^{(0)}_T T, 
	\label{PVT}
\end{equation} 
which differs from the usual thermal pressure of an ideal Bose gas by  the factor $h_{3}(\epsilon_{\text{qq}})/f_3(\epsilon_{\text{qq}})$ arising from the anisotropic QQI.

\section{Generalized uncertainty principle} \label{GUP}

The two most commonly used models of the GUP are : QGUP \cite{Kempf,Mag} and LQGUP \cite{Das, Ali}.

\subsection{QGUP} \label{LQGUP}

The QGUP relation which is inspired by string theory, it predicts the existence of a minimum measurable length 
$\Delta x_{\text{min}}= \hbar \sqrt{3 \beta}= \sqrt{3 \beta_0} l_p.$ \cite{Kempf}, 
where  $\beta >0$ is the deformation parameter which is often expressed in terms of the dimensionless parameter  $\beta_0$,
and  the Planck length $l_p$ as: $\beta= \beta_0 l_p^2/\hbar^2=\beta_0/(M_p c)^2$, where $M_p=\sqrt{\hbar c/G}$ is the Planck mass with
$G$ being the gravitational constant and $c$ denotes the speed of light in vacuum. 
The modified commutation relation reads
\begin{equation}\label {QGUP1}
[\hat x, \hat p]= i\hbar \left( 1+ \beta  p^2\right).
\end{equation}
Equation (\ref{QGUP1}) immediately leads to the uncertainty relation:
\begin{equation}\label {QGUP}
\Delta x \Delta p \geq  \frac{\hbar}{2}  \left[ 1+ \beta (\Delta p)^2\right].
\end{equation}
For $\beta=0$, the traditional Heisenberg principle (HUP) \cite{Heisenberg} is recovered. \\
The deformed density of states can be given by \cite{Kempf}
\begin{equation}\label {QGUP2}
{\cal D} (p) dp= \frac{V} {2 \pi^2 \hbar^3} \frac{p^2 dp} {(1+\beta p^2)^3}.
\end{equation} 
For $\beta \ll 1$, we expand the correction factor $(1+\beta p^2)^{-3}$ up to  second-order as: $(1+\beta p^2)^{-3}= 1-3\beta p^2 + 6( \beta p^2)^2 -\cdots$
Thus, the phase-space measure can be rewritten in term of energy as:
	\begin{align}\label {QGUP3}
	\sum_{\mathbf n} =\frac{V}{(2\pi \hbar)^3} \int_{0}^{\infty} d^3p 
= \frac{1}{2}\int_{0}^{\pi} \sin{\theta}d\theta \int_{0}^{\infty} {\cal D} (E)\,dE, 
	\end{align}
where
\begin{equation} \label{DoS}
{\cal D}(E)=\frac{(2m)^{3/2}}{4\pi^2 \hbar^3} E^{1/2} \left [1 - 6\,\beta m E+ 24 (\beta m E)^2 -\cdots\right].
\end{equation} 

\subsection{LQGUP} \label{LQGUP}

The LQGUP  is motivated by ideas  from doubly special relativity and phenomenology.  The commutation relation  becomes \cite{Ali}:
\begin{equation}\label {GUP}
[ r_i, p_j] = i\hbar  \left[ \delta_{ij}- \alpha \left( p \delta_{ij} + \frac{p_ip_j}{p}\right) + \beta \left(p^2 \delta_{ij} + 3 p_ip_j\right) \right],
\end{equation}
where  $x_i$ and $p_j$ are the position and momentum operators, respectively and $p=\sqrt{p_ip_j}$,
$\alpha= \alpha_0 l_p/\hbar=\alpha_0/(M_p c)$, and  $\beta= \beta_0 l_p^2/\hbar^2=\beta_0/(M_p c)^2$, where $\alpha_0$ and $\beta_0$ 
are the linear and quadratic deformation GUP parameters.
For $\alpha=\beta=0$, one can reproduce the standard HUP \cite{Heisenberg}.
The corresponding phase-space measure $ {\cal D} (E)$ is commonly written as \cite{Das}:
\begin{equation} \label{DoSLQ}
{\cal D}(E)=\frac{(2m)^{3/2}}{4\pi^2 \hbar^3} E^{1/2}(1+16\alpha\sqrt{m}E^{1/2}-25\beta m E).
\end{equation} 

We emphasize that the deformed commutation relation and the associated momentum-space Jacobian do not uniquely determine the Hamiltonian of the deformed theory. 
In the present work, we adopt  a specific quadratic-Hamiltonian realization (\ref{ham1}) in which the kinetic energy retains its standard nonrelativistc form $E_p=\ p^2/(2m)$,  and 
the quadrupolar interaction is specified by the Fourier-space potential given in  Eq.~(\ref{intP}),  while the QGUP/LQGUP deformation is introduced  through the density of state in (\ref{DoS}) and (\ref {DoSLQ}).
This choice is an  explicit modeling assumption defining the realization considered here, rather than a unique consequence of the deformed algebra. 
More general Hamiltonians containing deformation-dependent terms, such as \(O(\beta p^4)\), \(O(\beta^2p^6)\), or analogues LQGUP contributions,  and  interaction potential constructed utilizing a specific GUP-modified Fourier transform,
would lead to additional corrections to the thermodynamics quantities. Their coefficients cannot be determined from the deformed commutation relation alone and therefore  require additional microscopic input.  We do not introduce such undetermined terms here. Therefore all results  obtained below should be understood as predictions within the quadratic-Hamiltonian GUP/LQGUP realization specified alone.


The perturbation expansion of the modified momentum measure is valid provided that the GUP deformation remains small over the momentum range that contributes appreciably to the thermodynamic integrals. In particular, the expansion requires 
\begin{equation} \label{CC1}
\beta p^{2}\ll 1,
\qquad
\alpha p\ll 1.
\end{equation} 
Since the thermodynamic integrals involve a continuous range of momenta, these conditions should be evaluated at the characteristic momentum of the thermally populated quasiparticles,  rather than at an arbitrary momentum. 
Therefore, the perturbative GUP expansion is controlled when
\begin{equation} \label{CC2}
2m\beta T\ll1,
\qquad
\alpha\sqrt{2mT}\ll1.
\end{equation} 
For the LQGUP calculation, it is truncated  at first order in \(\alpha\) and \(\beta\).

To ensure  that the truncation of the QGUP/LQGUP expansion is self-consistent,  conditions (\ref{CC1}) and (\ref{CC2})  must be fulfilled  for each temperature and parameter set used in the numerical results.

\section{ Thermostatistics of a quadrupolar BEC under the GUP}

In this section we use the above phase-space approach of GUP, in which the Bogoliubov excitation spectrum of a quadrupolar BEC remains unchanged, while the density of states is modified
through  the replacements of integration measures described by Eqs.~(\ref{DoS}) and (\ref {DoSLQ}), enabling us to use thermodynamics as a tool to explain QG effects.

Starting from the GUP-modified Bogoliubov equations, we derive the thermal contribution to the condensate depletion,  condensed fraction,  the critical temperature, 
the grand potential via the partition function, and then obtain the corresponding EoS. Other thermodynamics quantities such as the internal energy, the heat capacity, and the entropy 
are calculated in Appendix B.

\subsection{Condensed fraction}

In the realm of the GUP, the high-momentum thermal depletion can be expressed as:
\begin{align}\label {norTGUP}
\tilde{n}_T&= \frac{1}{4}\int_{0}^{\pi} \sin{\theta}d\theta \int_{0}^{\infty} {\cal D} (E)\, \left[A (\theta)+A^{-1} (\theta)\right]  \frac{dE}{\exp(A (\theta) E/T)-1}.  
\end{align}
Using Eq.~(\ref{DoS}), we obtain the QGUP-corrected thermal depletion:
\begin{align}\label {norQGUP}
\tilde{n}_T^{\text{QGUP}}= 
 \tilde{n}_T^{(0)}\bigg[1-  9 \beta m T \frac{\zeta(5/2)}{\zeta(3/2)} \frac{f_5(\epsilon_{\text{qq}})}{f_3(\epsilon_{\text{qq}})} 
+ \frac{\zeta(7/2)}{\zeta(3/2)} 90 (\beta m T)^2 \frac{f_7(\epsilon_{\text{qq}})}{f_3(\epsilon_{\text{qq}})}\bigg]. 
\end{align}
Near the transition $T\rightarrow T_c$, the chemical potential decreases with temperature and the thermal depletion becomes comparable with the total density, $n_T \approx n$. 
One should stress also that it is hard to extract a closed form expression of \(T_c\) from Eq.~(\ref{norQGUP}). Therefore, a perturbative approach is highly needed \cite{Das}.
One defines 
\[
T_c ^{\text{QGUP}}= T_c^{(0)} + \Delta T (\beta) + \Delta T (\beta^2),
\]
where  \(T_c^{(0)}\) is the critical temperature without GUP derived from Eq. (\ref{nth})
\begin{equation}
	T_c^0 = \frac{2 \pi \hbar^2}{m }   \bigg[ \frac{4\,n}{  \zeta(3/2) f_3 (\epsilon_{\text{qq}}) } \bigg]^{2/3}.
\end{equation}
Then, the QG-corrected critical temperature reads
\begin{equation} \label{Tc-model1}
T_c^{\mathrm{QGUP}}=T_c^{(0)}
\left\{
1
+
6
\frac{\zeta(5/2)}{\zeta(3/2)}
\frac{f_5(\epsilon_{qq})}{f_3(\epsilon_{qq})}
\beta m T_c^{(0)}
+
\left[
81
\left(
\frac{\zeta(5/2)}{\zeta(3/2)}
\frac{f_5(\epsilon_{qq})}{f_3(\epsilon_{qq})}
\right)^2
-
60
\frac{\zeta(7/2)}{\zeta(3/2)}
\frac{f_7(\epsilon_{qq})}{f_3(\epsilon_{qq})}
\right]
\left(\beta m T_c^{(0)}\right)^2
\right\},
\end{equation}
which shows that the QG corrections increase with increasing density, \(n\), and decreasing boson mass, \(m\). 
One can also see that higher-order QG corrections (\(\propto \beta^2\)) have a stronger dependence on \(n\).  

The magnitude of the relative temperature shift, $\Delta T= T_c-T_c^{(0)}$, can be easily extracted as:
\begin{equation}
\frac{\Delta T^{\text{QGUP}}}{T_c^{(0)}}= 
6
\frac{\zeta(5/2)}{\zeta(3/2)}
\frac{f_5(\epsilon_{\rm qq})}{f_3(\epsilon_{\rm qq})}
\beta m T_c^{(0)}
+
\left[
81
\left(
\frac{\zeta(5/2)}{\zeta(3/2)}
\frac{f_5(\epsilon_{\rm qq})}{f_3(\epsilon_{\rm qq})}
\right)^2
-
60
\frac{\zeta(7/2)}{\zeta(3/2)}
\frac{f_7(\epsilon_{\rm qq})}{f_3(\epsilon_{\rm qq})}
\right]
\left(\beta m T_c^{(0)}\right)^2.
\end{equation} 
For a stable  quadrupolar  gas, $f_3(\epsilon_{\rm qq})>0$, $f_5(\epsilon_{\rm qq})>0$ and  for a positive $\beta$, one has $\Delta T>0$ at leading order.
An  important remark is that the second order term does not reverse the leading sign (partially compensate it) with the perturbative regime unless  \(\beta mT_c^{(0)}\) becomes sufficiently large such as the perturbative expansion itself becomes no longer valid.

The condensed fraction is defined as $n_c/n=1-\tilde n/n$.  Including corrections up to second order in $\beta$, we obtain
\begin{equation} \label{CF1}
\frac{n_c^{\mathrm{QGUP}}}{n}
=
1-
\left(\frac{T}{T_c^{(0)}}\right)^{3/2}
\left[
1
-
9\beta m T_c^{(0)}
\frac{\zeta(5/2)}{\zeta(3/2)}
\frac{f_5(\epsilon_{\rm qq})}{f_3(\epsilon_{\rm qq})}
\left(\frac{T}{T_c^{(0)}}\right)
+
90(\beta m T_c^{(0)})^2
\frac{\zeta(7/2)}{\zeta(3/2)}
\frac{f_7(\epsilon_{\rm qq})}{f_3(\epsilon_{\rm qq})}
\left(\frac{T}{T_c^{(0)}}\right)^2
\right].
\end{equation}

Following the same fashion and using Eq.~(\ref{DoSLQ}), we obtain the LQGUP corrected thermal depletion:
\begin{align}\label {norLGGUP}
\tilde n_T^{\text{LQGUP}}=\tilde{n}_T^{(0)}\bigg[1+  \frac{16 \pi^{3/2}}{3 \zeta(3/2)} \frac{f_2(\epsilon_{\text{qq}})}{f_3(\epsilon_{\text{qq}})}  \alpha \left(m T\right)^{1/2} 
- \frac{75\zeta(5/2)}{2\zeta(3/2)} \frac{f_5(\epsilon_{\text{qq}})}{f_3(\epsilon_{\text{qq}})}  (\beta m T) \bigg]. 
\end{align}
Using the perturbed relation, $T_c = T_c^{(0)} + \Delta T(\alpha) + \Delta T(\beta)$, for the critical temperature, we find
\begin{equation} \label{Tc-model2}
T_c^{\text{LQGUP}}=T_c^0 \bigg[  1 -  \frac{32  \pi^{3/2} f_2(\epsilon_{\text{qq}})}{9 \zeta(3/2) f_3 (\epsilon_{\text{qq}})}   \alpha \left(m T_c^{0}\right)^{1/2}  
+ 25  \frac{ \zeta(5/2) f_5(\epsilon_{\text{qq}})}{ \zeta(3/2)  f_3 (\epsilon_{\text{qq}})} (\beta m T_c^{0}) \bigg].
\end{equation}
Thus, the magnitude of the relative temperature correction reads
\begin{equation} \label{LQGUPTc}
\frac{\Delta T^{\text{LQGUP}}}{T_c^{(0)}} = -  \frac{32  \pi^{3/2} f_2(\epsilon_{\text{qq}})}{9 \zeta(3/2) f_3 (\epsilon_{\text{qq}})}   \alpha \left(m T_c^{0}\right)^{1/2}  
+ 25  \frac{ \zeta(5/2) f_5(\epsilon_{\text{qq}})}{ \zeta(3/2)  f_3 (\epsilon_{\text{qq}})} (\beta m T_c^{0}).
\end{equation}
This clearly indicates that  $\Delta T_c^{(0)} (\alpha)  < 0 $  and $ \Delta T_c^{(0)} (\beta)> 0$. So, $\alpha$ tends  to increase $\tilde n_T$ and decrease $T_c$ while $\beta$  has opposite effects.
One can also show that  the two LQGUP contributions exhibit competing signs with density  $\propto -\alpha n^{1/3}+ \beta n^{2/3}$.

Incorporating both the linear, $\alpha$, and the quadratic, $\beta$, GUP deformation parameters yields the following LQGUP  condensed fraction:
\begin{equation} \label{NC}
\begin{aligned}
\frac{n_c^{\text{LQGUP}}}{n} = 1 - \left(\frac{T}{T_c^{(0)}}\right)^{3/2} - \frac{16 \pi^{3/2}}{3 \zeta(3/2)} \frac{f_2(\epsilon_{\rm qq})}{f_3(\epsilon_{\rm qq})} \alpha \sqrt{m T_c^{(0)}} \left(\frac{T}{T_c^{(0)}}\right)^2 + \frac{75 \zeta(5/2)}{2 \zeta(3/2)} \frac{f_5(\epsilon_{\rm qq})}{f_3(\epsilon_{\rm qq})} \left(\beta m T_c^{(0)}\right) \left(\frac{T}{T_c^{(0)}}\right)^{5/2},
\end{aligned}
\end{equation}

Our results show that the long-range anisotropic QQI modifies the Bogoliubov spectrum, which increases thermal depletion even without QG corrections. 
For  $\epsilon_{\text{qq}} \ll 1$,  Eqs.~(\ref{norQGUP})-(\ref{NC}) reduce to those of a weakly-interacting BEC with a pure contact interaction. 
In the absence of both QQI and contact interaction one recovers the results of ideal gases under the QGUP \cite{VaK,Zhang, Cast, Li} and LQGUP \cite{Das}.
When $\alpha= \beta =0$, we reproduce our Eqs.~(\ref{nth})-(\ref{PVT}).

It is also important to clarify the validity of the obtained transition temperature. The Bogoliubov approximation is formulated around a macroscopically occupied condensate and therefore becomes progressively less controlled as the condensate density  decreases on approaching the transition. In the present work, the condition ($\tilde n_T \simeq n$) is used as a criterion for estimating the temperature at which the thermal population becomes comparable to the total density with the adopted high-momentum Bogoliubov framework. Therefore, the resulting  $T_c$ should be regarded as a Bogoliubov-level transition-temperature estimate rather than as an exact determination of the critical point of the interacting system.  A quantitative precise description arbitrarily close to the transition would require a self-consistent finite-temperature treatment of the condensate and noncondensate components together with the inclusion of long-wavelength critical fluctuations. 
The GUP-induced shifts obtained below above should consequently interpreted in the regime where the condensate remains sufficiently populated, the high-momentum approximation is applicable, and the GUP expansion remains perturbative. 
Within these conditions, the calculation is consistently isolates the modification of the thermal phase-space measure produced by both QGUP and LQGUP and determines its corresponding effect on $T_c$.

\begin{figure}
\centering 
\includegraphics[scale=0.7, angle=0] {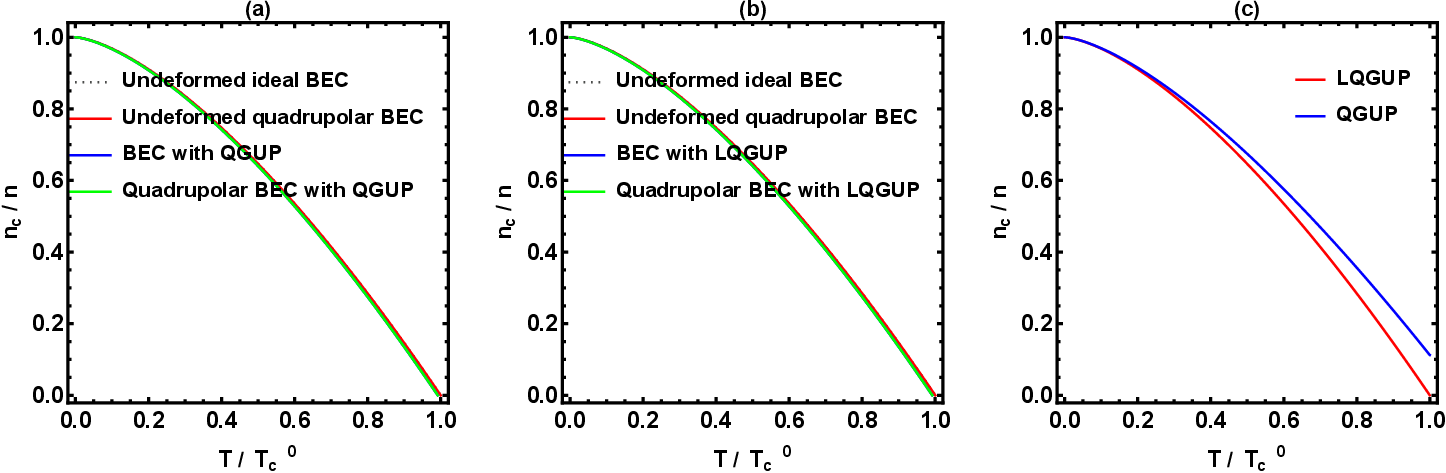}
\caption {Condensed fraction, $n_c/n$, versus critical temperature, $T/T_c^{(0)}$, for $\epsilon_{\text{qq}}=0.01$ and $\beta_0= 2.47 \times 10^{46}$. 
(a) QGUP model. (b) LQGUP model. (c) Comparison between QGUP and LQGUP models.}
\label{F1}
\end{figure}

To gain further insights into the behavior of the condensed fraction, $n_c/n$, we plot it as a function of $T/T_c^{(0)}$.
As shown in Fig.~\ref{F1}, the anisotropic  long-range QQI modifies the quasiparticle spectrum and enhances the thermal depletion giving rise to a reduction of the condensate fraction.
Therefore, the thermal fluctuation becomes strong enough to destroy the condensate even at temperatures well below ideal-gas critical temperature, $T_c^{(0)}$, as  anticipated above  in Eqs.~(\ref{norQGUP}) and  (\ref {norLGGUP}).
The inclusion of the LQGUP furnishes additional QG corrections through the modified phase-space measure,  inducing a larger reduction in $n_c/n$  than the QGUP notably at higher temperatures for the parameter choices considered here
 (see Fig.~\ref{F1} (c)).

\begin{figure}
\centering 
\includegraphics[scale=0.5, angle=0] {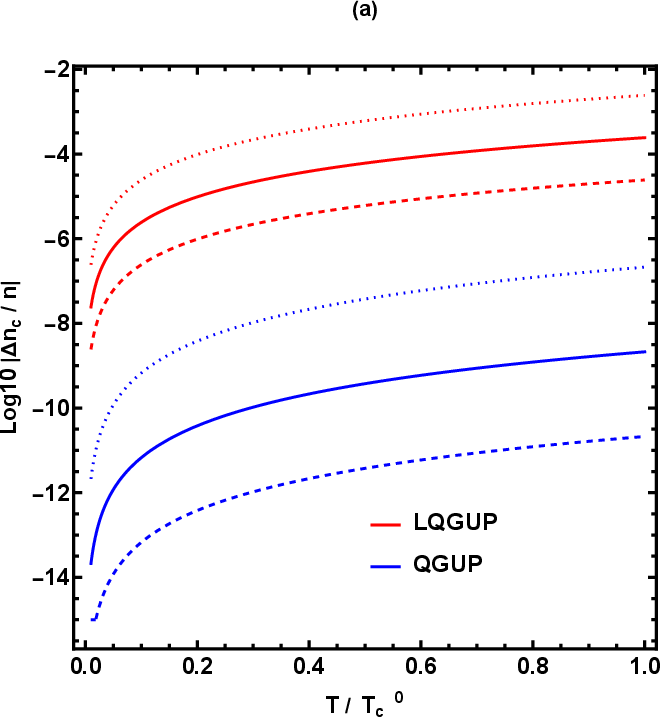}
\includegraphics[scale=0.5, angle=0] {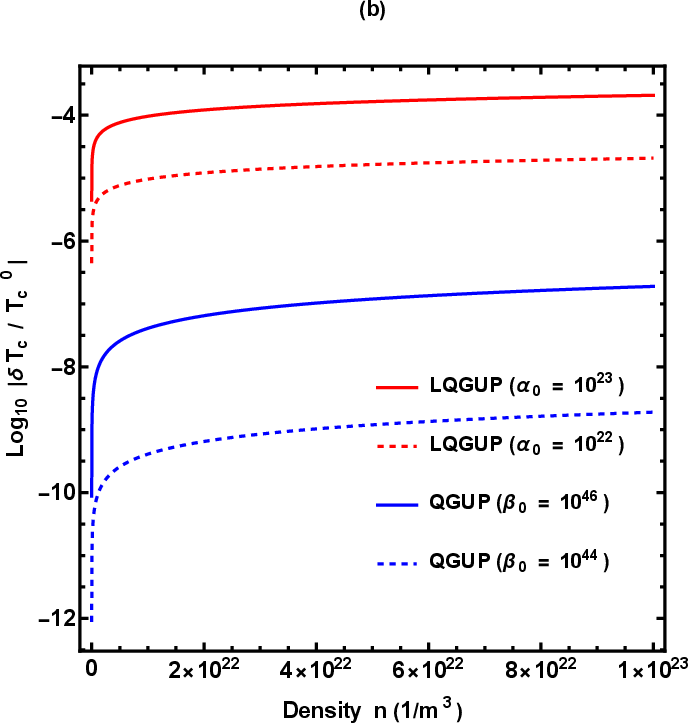}
\caption { (a) Relative condensed fraction correction, $\text{Log}_{10} |\Delta n_c/n|$, induced by QGUP and LQGUP, versus critical temperature, $T/T_c$ for different values of  the GUP parameters. 
Here we set $\alpha = \sqrt{\beta}$ \cite{Das}.  Solid lines correspond to $\beta_0= 10^{46}$. Dashed lines  correspond to $\beta_0= 10^{44}$. Dotted lines correspond to $\beta_0= 10^{48}$.
(b) Relative temperature correction, $\text{Log}_{10}|\Delta T/T_c^{(0)}|$, versus the density, $n$, for different values of  the GUP parameters with $\epsilon_{\text{qq}}=0.01$. }
\label{F2}
\end{figure}

Figure ~\ref{F2}  (a) compares $\text{Log}_{10} |\Delta n_c/n|$ induced by QGUP and LQGUP for several deformation parameters. 
We see that  the LQGUP predicts a stronger shift.  Here, a larger curve for a particular parameter choice does not prove that LQGUP universality produces a larger effect. 
The comparison depends on the values and units assigned to $\alpha_0$ and $\beta_0$, and on whether the same perturbative regime is respected. 
Increasing the deformation parameter rise the magnitude of the correction, as expected within the expansion. 
It is worth noticing that  the condensed fraction shift not only originate from the GUP effects but also from the quadrupolar interaction.



The density dependence of the relative critical temperature shift, $\text{Log}_{10}|\Delta T/T_c^{(0)}|$, serves as a key thermodynamic observable for distinguishing between  QGUP and  LQGUP corrections in quadrupolar BECs.
The structural difference between linear and quadratic Planck-scale momentum deformation is directly imprinted on the macroscopic density power laws (see Sec.~\ref{bound}  for details).
The LQGUP corrections scale with the inverse mean interparticle distance, $n^{1/3} \sim 1/d$. 
 As shown in Fig.~\ref{F2} (b) When $\log_{10} |\Delta T_c / T_c^{(0)}|$ is evaluated against \(n\)  $n$ on a standard linear axis, the $n^{1/3}$ term exhibits a steep, concave initial rise at moderate gas densities  ($n < 2 \times 10^{22} \text{ m}^{-3}$) 
 before approaching sublinear  regime. This indicates that the LQGUP effects provide a strongly enhanced sensitivity profile in low-to-intermediate density regimes.
However,  the QGUP scales as  $n^{2/3} \sim 1/d^2$, while also growing sublinearly relative to  $n$, $n^{2/3}$ climbs noticeably faster than $n^{1/3}$ at higher limits ($n \to 10^{23} \text{ m}^{-3}$) as seen  in Fig.~\ref{F2} (b).

In the combined LQGUP scenario where both linear and quadratic terms compete, the zero-shift line  ($\Delta T_c = 0$) defines a degenerate parameter trajectory  $\beta_0 / \alpha_0 = \chi_\alpha(n) / \chi_\beta(n) \propto n^{-1/3}$. 
Performing thermometry across multiple density operational point breaks this parameter degeneracy and isolates the individual linear ($\alpha_0$)  and quadratic  ($\beta_0$) coupling bounds.



\subsection{Equation of state}

The GUP  high-momentum corrected thermal EoS is given by:
\begin{equation}
\frac{P_TV}{T}=-\frac{1}{2} \int_{0}^{\pi} \sin{\theta}d\theta \int_{0}^{\infty} {\cal D} (E) \ln \left(1- e^{-A (\theta) E/T}\right) dE.
\end{equation} 

The QGUP  deformed measure of Eq.~(\ref{DoS}) yields:
\begin{equation} \label{thPQ}
\mathcal P^{\text{QGUP}}= \left(\frac{P_TV}{nT}\right)^{\text{QGUP}}
=
\frac{2\zeta(5/2)}{\zeta(3/2)}\frac{h_3(\epsilon_{qq})}{f_3(\epsilon_{qq})}
\left(\frac{T} {T_c^{(0)}} \right)^{3/2}
+
\Delta P^{\rm QGUP},
\end{equation}

Upon consistently expressing the result in terms of normalized temperature, $T/T_c^{(0)}$,  we  obtain for the relative QGUP correction to the  thermal  part of the EoS.
\begin{equation} \label{thPQ}
\Delta \mathcal{P}^{\text{QGUP}} = - 18 (\beta m T_c^{(0)}) \frac{\zeta(7/2)}{\zeta(3/2)} \frac{h_5(\epsilon_{qq})}{f_3(\epsilon_{qq})} \left(\frac{T}{T_c^{(0)}}\right)^{5/2} + 180 (\beta m T_c^{(0)})^2 \frac{\zeta(9/2)}{\zeta(3/2)} 
\frac{h_7(\epsilon_{qq})}{f_3(\epsilon_{qq})} \left(\frac{T}{T_c^{(0)}}\right)^{7/2}. 
\end{equation}
The second-order terms are appealing  since  they contain not only the explicit $\beta^2$ contribution from the density of state, but also the  $\beta$-dependence generated when  $T_c^{\text{QGUP}}$ is used to normalize $T$.

\begin{figure}
\centering 
\includegraphics[scale=0.7, angle=0] {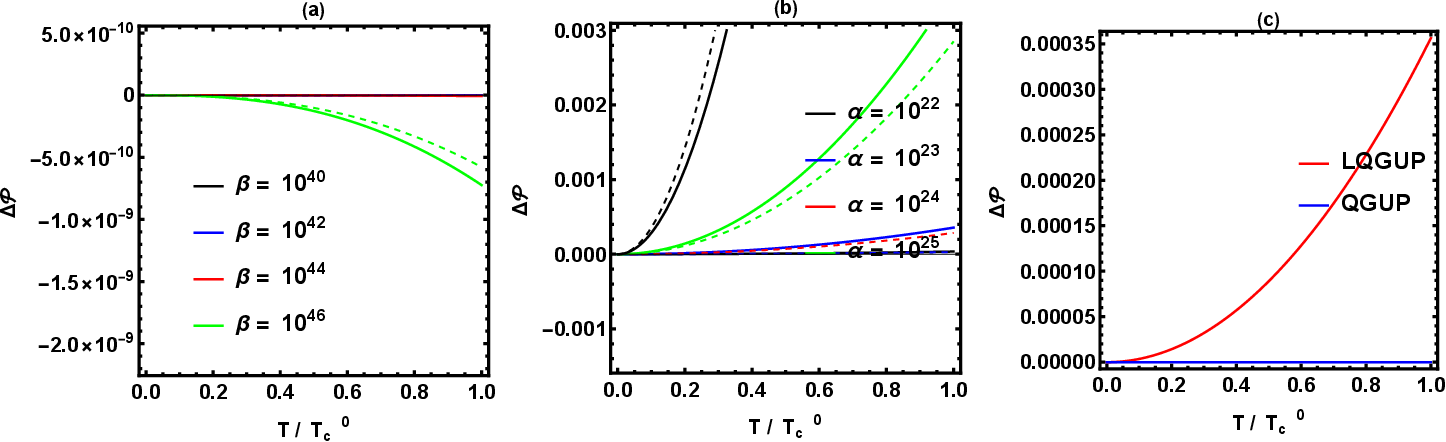}
\caption { Thermal EoS correction, $\Delta \mathcal P$, versus critical temperature, $T/T_c$, for different values of  the GUP parameters with $\epsilon_{\text{qq}}=0.01$. 
(a) QGUP model. (b) LQGUP model. (c) Comparison between QGUP and LQGUP models.
Dashed lines corresponds to nonquadrupolar BECs, $\epsilon_{\text{qq}}=0$.}
\label{F3}
\end{figure}

Applying now the LQGUP modified density of states and performing the energy and angular integrations one finds  for the LQGUP thermal pressure: 
\begin{equation}
\mathcal P^{\rm LQGUP}
=
\frac{2\zeta(5/2)}{\zeta(3/2)}
\frac{h_3(\epsilon_{qq})}{f_3(\epsilon_{qq})}
\left(\frac{T}{ T_c^{(0)}}\right)^{3/2}
+
\Delta P^{\rm LQGUP}.
\end{equation}

The normalized direct $T/T_c^{(0)}$ results yields for  the relative  LQGUP correction to the  thermal quasiparticle pressure up  to first order in  $\alpha$ and $\beta$
\begin{equation}  \label{thPLQ}
\Delta \mathcal{P}^{\text{LQGUP}}= \alpha \sqrt{m T_c^{(0)}} \frac{64\sqrt{\pi}\zeta(3)}{\zeta(3/2)} \frac{h_4(\epsilon_{qq})}{f_3(\epsilon_{qq})} \left(\frac{T}{T_c^{(0)}}\right)^2 - \beta m T_c^{(0)} \frac{75\zeta(7/2)}{\zeta(3/2)} \frac{h_5(\epsilon_{qq})}{f_3(\epsilon_{qq})} \left(\frac{T}{T_c^{(0)}}\right)^{5/2}.
\end{equation}
The leading term $\propto T^{3/2}$ is the standard thermal pressure of an ideal gas, subleading term $\propto \alpha\,T^{2}$ stands for the first-order correction, while the last term 
$\propto \beta\,T^{5/2}$ represents the quadratic GUP correction.

The behavior of  the corrected-thermal quasiparticle pressure $\Delta \mathcal P$ obtained from both GUP proposals is shown in Fig.~\ref{F3}. 
For the QGUP model, Eq.~(\ref{thPQ}) predicts a negative leading-order correction for positive $\beta$, while the positive second-order term offsets this reduction. The correction remains negative over the parameter and temperature ranges displayed in 
Fig.~\ref{F3} (a).  In contrast, the LQGUP correction in  Eq.~(\ref{thPLQ})  contains competing  contributions proportional to $\alpha$ and  $-\beta$. Its sign therefore depends on the relative magnitudes:  the correction is positive when the linear contribution dominates and negative when the quadratic contribution dominates.  The positive curves in Fig.~\ref{F3} (b) indicate that the linear contribution dominates for the parameter choices considered here. Accordingly, Fig.~\ref{F3} (c) illustrates a larger positive LQGUP correction than the QGUP effects over the displayed range, but this ordering is not universal and depends on the deformation parameters and on the validity of the perturbative expansion.

\section{Constraints on the GUP deformation parameters} \label{bound}

The modification of the critical  temperature provides a direct way of constraining the GUP deformation parameters from precision measurements of $T_c$
We characterize the experimental sensitivity by requiring that the GUP-induced shift remains smaller than the relative experimental uncertainty. 
\begin{equation} 
\left| \frac{\Delta T_c}{T_c^{(0)}} \right| < \delta , 
\label{eq:experimental_criterion}
 \end{equation} 
where $\delta$  is the attainable relative precision in the determination of the transition temperature.
The deformation parameters are related to the dimensionless parameters conventionally used in GUP phenomenology through
$$
 \alpha=\alpha_0\frac{\ell_{\rm P}}{\hbar}, \qquad \beta=\beta_0\frac{\ell_{\rm P}^2}{\hbar^2}. 
 \label{eq:GUP_dimensionless} 
$$
It is important to distinguish individual constraints from a simultaneous constraint on the two GUP parameters. In the LQGUP case, Eq.~\eqref{LQGUPTc}  has both  $\alpha$- and $\beta$-dependent contributions with opposite signs.
As a result, a single measurement of $T_c$ does not, in general, independently determine both parameters since partial cancellation between the two contributions is possible.  
We therefore adopt a one-parameter-at-a-time procedure.
For the constraint on $\alpha_0$, we set $\beta_0=0$, whereas for the constraint on $\beta_0=0$, we set  $\alpha_0=0$. 
The resulting limits should hence be interpreted as individual sensitivities of the transition temperature to the corresponding deformation parameter, rather than as a joint confidence region in the $(\alpha_0,\beta_0)$ plane. 

For the QGUP, retaining the leading contribution in  $\beta$, the relative shift can be written as:
 \begin{equation} 
 \frac{\Delta T_c^{\rm QGUP}}{T_c^{(0)}} = \chi_{\rm QGUP}\,\beta_0 +\mathcal{O}\!\left(\beta_0^2\right), \label{eq:CQGUP_definition} 
 \end{equation} 
where
$$
 \chi_{\rm QGUP} = \frac{3\pi\zeta(5/2)} {\zeta(3/2)^{5/3}} \frac{f_5(\epsilon_{qq})} {f_3(\epsilon_{qq})^{5/3}} \,\ell_{\rm P}^2 n^{2/3}. 
$$
Therefore, the corresponding constraint reads  
 \begin{equation} 
 \beta_0 < \frac{\delta}{\chi_{\rm QGUP}}. 
 \label{eq:QGUP_bound} 
 \end{equation} 
 For the representative parameters  $ n=10^{20}\,{\rm m}^{-3}$, $\epsilon_{\rm qq}=0.01$,   we obtain  $\chi_{\rm QGUP} \simeq 5.7\times10^{-57}$ .
Thus, a relative precision of $\delta=10^{-3}$,  gives
  \begin{equation} 
 \beta_0^{\rm QGUP} \lesssim 1.75 \times10^{53} . 
  \label{eq:QGUP_numerical_bound} 
  \end{equation} 
In the case of the LQGUP, Eq.~\eqref{LQGUPTc} can be written in the form 
  \begin{equation} 
  \frac{\Delta T_c^{\rm LQGUP}}{T_c^{(0)}} = -\chi_{\alpha}\alpha_0 + \chi_{\beta}\beta_0, \label{eq:LQGUP_coefficients} 
  \end{equation} 
  where
$$ 
  \chi_{\alpha} = \frac{32\sqrt{2}\pi^2} {9\zeta(3/2)^{4/3}} \frac{f_2(\epsilon_{qq})} {f_3(\epsilon_{qq})^{4/3}} \,\ell_{\rm P}n^{1/3}, 
$$
and 
$$
  \chi_{\beta} = \frac{100\pi\zeta(5/2)} {\zeta(3/2)^{5/3}} \frac{f_5(\epsilon_{qq})} {f_3(\epsilon_{qq})^{5/3}} \,\ell_{\rm P}^2n^{2/3}.
  $$ 
Setting $\beta_0=0$  gives the one-parameter constraint
 \begin{equation} 
\alpha_0 < \frac{\delta}{\chi_{\alpha}}, 
\label{eq:alpha_bound}
 \end{equation} 
 whereas setting $\alpha_0=0$ gives
  \begin{equation}
\beta_0 < \frac{\delta}{\chi_{\beta}}. 
 \label{eq:beta_bound} 
 \end{equation} 
 For the above benchmark parameters, the coefficients are  $ \chi_{\alpha} \simeq 6.5\times10^{-28}$, and  $\chi_{\beta} \simeq 1.9\times10^{-55}$. 
 Thus, for $\delta=10^{-3}$, 
 \begin{equation} 
\alpha_0^{\rm LQGUP} \lesssim 1.5\times10^{24}, \qquad  \beta_0^{\rm LQGUP} \lesssim 5.3\times10^{51}. 
 \label{eq:LQGUP_numerical_bounds} 
 \end{equation} 
 The above limits are one-parameters-at-a-time constraints and should not be interpreted as simultaneous bounds on $\alpha_0$ and $\beta_0$. In particular, because the two terms in Eq.~\eqref{eq:LQGUP_coefficients} have opposite signs, allowing both parameters to vary simultaneously would permit partial cancellation. 
 A statistically meaningful joint constraint would therefore require either additional independent observables or a dedicated two-parameters likelihood analysis. 
 One may alternatively impose a model-dependent relation between the two deformation parameters. A commonly used choice is 
 \begin{equation} 
 \alpha_0=\sqrt{\beta_0}. 
 \label{eq:alpha_beta_relation}
  \end{equation} 
  In this case Eq.~\eqref{eq:LQGUP_coefficients} becomes a one-parameter expression and a corresponding constraint can be obtained. For the present benchmark, this gives 
 \begin{equation} 
\beta_0\lesssim 1.2 \times10^{52}, \qquad \alpha_0\lesssim 1.1 \times10^{26}  \qquad \left(\alpha_0=\sqrt{\beta_0}\right). 
 \label{eq:correlated_bound} 
 \end{equation} 
 These values are not independent constraints on  $\alpha_0$ and $\beta_0$; they follow only after imposing  the model in  Eq.~\eqref{eq:alpha_beta_relation}. 
 Similarly for  $\delta = 10^{-4}$  gives  $\beta_0^{\rm QGUP} \lesssim 1.75 \times10^{52}$, and   $\alpha_0^{\rm LQGUP} \lesssim 1.5\times10^{23}$,  $\beta_0^{\rm LQGUP} \lesssim 5.3\times10^{50}$. 
 Imposing additionally  the model relation $\alpha_0=\sqrt{\beta_0}$,  we get  $\beta_0 \lesssim 5.3\times10^{50}$,  and $ \alpha_0\lesssim 2.5 \times10^{25} $. 
 Therefore, quadrupolar BCE precision thermometry offers orders-of-magnitude tighter constraints on linear GUP parameters.

Representative constraints reported in the literature include  $\beta_0\lesssim10^{21}$ from scanning-tunnelling-microscope measurements \cite{Das11} ,  $\beta_0\lesssim10^{36}$ from the Lamb shift \cite{Das11} , 
 $\beta_0\lesssim10^{50}$  from Landau-level  measurement \cite{Das11},  $\alpha_0 \lesssim 10^{19}$ and $\beta_0 \lesssim 10^{46}$ statistical mechanics \cite{Das},  
 $\alpha_0 \lesssim 10^{14}$, $\beta_0\lesssim10^{39}$ using atom-recoil measurements in $^{87}$Rb   \cite{Gao}.
In our recent study based on the Hartree-Fock-Bogliubov theory \cite{Boudj, Boudj2}, we  have demonstrated that weakly interacting BEC with contact and dipolar interactions provide improved
limits on the GUP parameters. Using realistic $^{52}$Cr and $^{168}$Er condensates, we found $\alpha_0 \lesssim 2.6-3.4 \times 10^{22}$ and $\beta_0 \lesssim  6.7 \times 10^{44}-1.2 \times 10^{45}$ 
from the condensate fraction, while the superfluid fraction gave  $\alpha_0 \lesssim 10^{24}$ and $\beta_0 \lesssim 10^{48}$ \cite{Boudj2}. 
The major discrepancy between these dipolar bounds and the present quadrupolar ones points out that the two analyses are not employing the same observables, normalization, and extraction procedure. 
A new formulation of the GUP based on quantum droplets of binary BECs yields also an ameliorated bounds $\beta_0 \lesssim 10^{49}$ \cite{Boudj1}.
The projected sensitivities obtained here are therefore not the most stringent existing constraints on the GUP parameters. Rather, they provide independent and complementary sensitivities associated with qualitatively different observables, namely, 
the finite-temperature transition  of a long-range quadrupolar quantum fluid. 
The sensitivity of the proposed measurement scales with the density according to $ \chi_{\alpha}\propto n^{1/3}$,  and  $\chi_{\beta}\propto n^{2/3}$  while the quadrupolar contribution enters through the functions $f_j(\epsilon_{\rm qq})$. 
Consequently, improved transition-temperature precision, higher molecular densities, and a quantitatively established quadrupole moment for the molecular state can substantially improve the projected sensitivity.

Although a BEC with electric QQIs  has not yet been experimentally realized, theoretical studies have demonstrated that quadrupolar  interaction of experimentally relevant magnitude can be incorporated into ultracold Bose gases \cite{Boudj,Wang, Zheng}. 
Recent observations of BEC in ultracold polar molecules further establish the experimental feasibility of molecular quantum gases with controllable long-range interaction  \cite{Biga, Shi}.
These developments provide a realistic pathway toward the quadrupolar molecular condensate considered here. 
The present analysis demonstrates that precision measurements of the critical temperature can provide an additional low-energy channel for testing GUP-induced modifications of phase-space counting.

\begin{table*}
\centering
 \caption{Illustrative projected QGUP and LQGUP sensitivity parameters for candidate homonuclear dimers. The estimates assume $T_c^{(0)}=300\,\mathrm{nK}$, a particle density $n=10^{20}\,\mathrm{m}^{-3}$, weak quadrupolar coupling, and a fractional temperature sensitivity $\delta=10^{-4}$. The molecular masses are used for illustrative mass rescaling at fixed critical temperature. These values are projections, not molecule-specific experimental constraints.} 
 \label{tab:molecular_GUP_bounds} 
 \begin{tabular}{lcccc} 
 \hline\hline Molecule & Mass (u) & $\beta_{0,\mathrm{QGUP}}^{\mathrm{proj}}$ & $\alpha_{0,\mathrm{LQGUP}}^{\mathrm{proj}}$ & $\beta_{0,\mathrm{LQGUP}}^{\mathrm{proj}}$ \\ \hline ${}^{52}\mathrm{Cr}_2$ & 104 & $1.93\times10^{51}$ & $1.02\times10^{22}$ & $4.64\times10^{49}$ \\ ${}^{84}\mathrm{Sr}_2$ & 168 & $1.20\times10^{51}$ & $8.01\times10^{21}$ & $2.87\times10^{49}$ \\ ${}^{133}\mathrm{Cs}_2$ & 266 & $7.55\times10^{50}$ & $6.36\times10^{21}$ & $1.81\times10^{49}$ \\ ${}^{174}\mathrm{Yb}_2$ & 348 & $5.78\times10^{50}$ & $5.57\times10^{21}$ & $1.39\times10^{49}$ 
 \\ \hline\hline 
 \end{tabular}
  \end{table*}
 The estimated bounds  presented in Table~\ref{tab:molecular_GUP_bounds}  clearly show that the mass dependence alone may change the projected GUP sensitivity under the stated assumptions.

\begin{figure}
\centering 
\includegraphics[scale=0.7, angle=0] {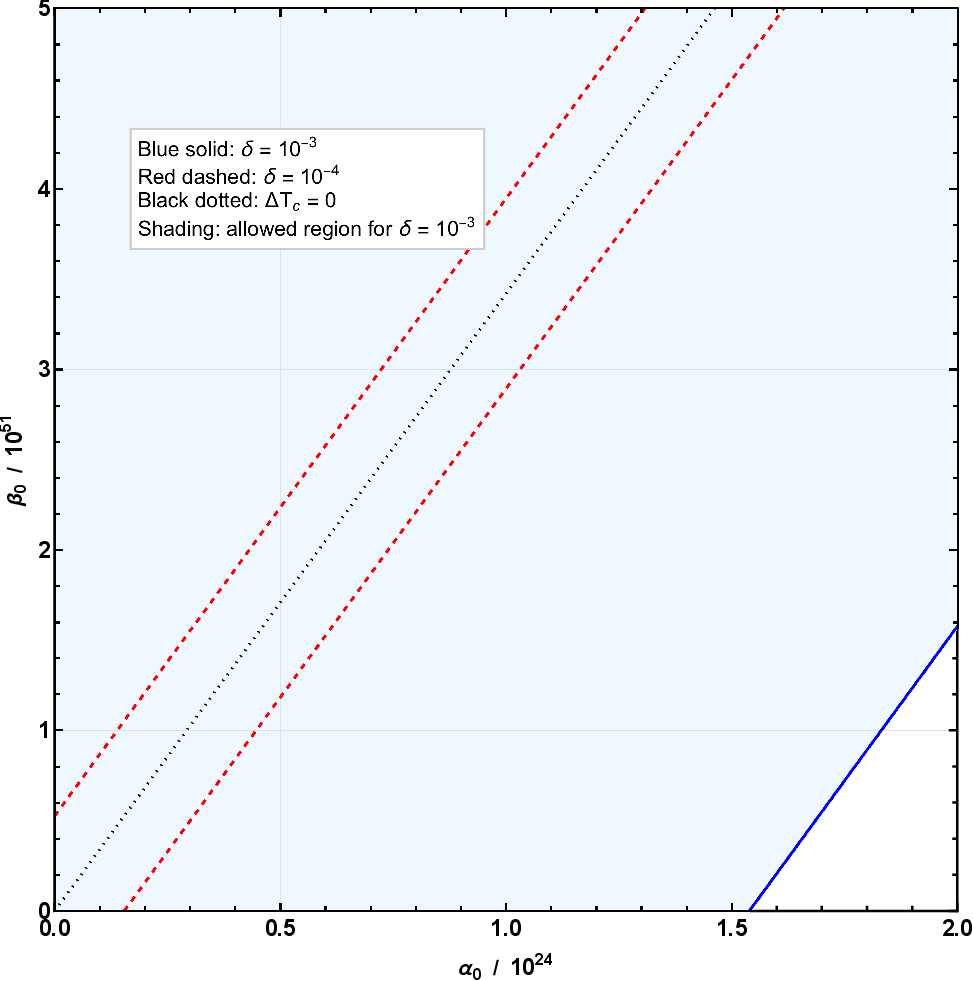}
\caption { Projected bounds contour region  $\alpha_0$  vs. $\beta_0$  from critical-temperature precision.}
\label{FF4}
\end{figure}

Figure \ref{FF4}   depicts the projected bounds on the GUP deformation parameters  ($\alpha_0, \beta_0$) derived from precision measurements of the critical temperature shift in a quadrupolar fluid.
The contour equation $-\chi_{\alpha}  \alpha_0 + \chi_{\beta} \beta_0= \delta$ describes a straight line with slope $d \beta_0/ d\alpha_0=   \chi_{\alpha}  /  \chi_{\beta}  \approx 3.4 \times 10^{27}$.
Since  $ \chi_{\alpha} $ and  $\chi_{\beta}$ have opposite signs in the LQGUP critical temperature shift, there exists a degenerate direction along $\beta \approx  (\chi_{\alpha}  /  \chi_{\beta} ) \alpha_0$, where the linear and quadratic QG corrections partially cancel each other out.
Such a cancellation  explains why setting $ \alpha_0=0$ yields a tight bound on $\beta_0  \leq 5.3 \times 10^{51}$, but allowing both parameters to vary simultaneously permits much larger individual values 
without violating the experimental precision threshold $\delta$.
Consequently, the critical-temperature measurements constrain a combination of $\alpha_0$ and $\beta_0$, rather than necessarily constraining each parameter independently. The cancellation line makes this limitation explicit.

\section{Conclusions} \label{concl} 

In this work we investigated the finite-temperature thermodynamics of homogeneous quadrupolar BECs  within  two commonly studied GUP prescriptions: the QGUP and LQGUP.
Our analysis  focused  on how the GUP-modified phase-space density of states affects the thermodynamic behavior of the system which is governed by a highly anisotropic quasiparticle spectrum arising from QQIs. 
We emphasize that we have adopted an effective phase-space approach in which the standard microscopic interaction Hamiltonian and the corresponding Bogoliubov excitation spectrum are retained, while momentum-space integration measure is modified according to the selected GUP adopted model.  This is a specific phenomenological prescription for investigating possible QG signatures in an interacting quantum gas; it is not a unique consequence of the GUP alone.  In particular, possible GUP-induced modifications of the kinetic operator, interaction matrix elements, or excitation spectrum would require an explicitly specified representation of the deformed position and momentum operators and a consistent derivation of the many-body Hamiltonian.

Within this framework we derived perturbative  expressions for the thermal depletion,  condensate fraction, critical temperature,  EoS,  internal energy,   heat capacity,  and entropy  in term of the system parameters.
 The validity criteria of the obtained analytical formulas are  accurately established.
The results show that the leading GUP corrections depend on both the deformation parameters and the anisotropic interaction coefficients. Consequently, the thermal response of quadrupolar condensate differs from that of an ideal Bose gas and depends on the strength and angular structure of the QQIs. The QGUP and LQGUP frameworks also  produce different temperature dependencies because their phase-space measures contain different powers of momentum. 
In the parameter regime considered here, the LQGUP can produce larger corrections than the QGUP. This comparison, however, is conditional on the chosen deformation parameters and on the validity of the perturbative expansion; it should not be interpreted as a universal ordering of the two models.  

We also examined how finite-temperature observables could be used to estimate sensitivity to the GUP deformation parameters.  The resulting estimates are projected sensitivities obtained by requiring the predicted temperature shift to remain below an assumed fractional resolution, $\delta_T$.  Moreover, in the LQGUP model, the linear and quadratic contributions can have opposite signs and may partially cancel.  Accordingly, bounds obtained by varying one deformation  parameter at a time should be distinguished from the joint allowed region in the   $(\alpha_0,\beta_0)$  plane. The frequently used relation $\alpha_0=\sqrt{\beta_0}$  represents a particular benchmark trajectory, rather than a universal requirement  of the LQGUP framework.
Indeed, a realistic assessment of  quadrupolar condensate as a probe of GUP-induced effects requires further work. In particular, candidate  molecular systems must be characterized using their state-specific quadrupole moments, scattering lengths, densities and stability conditions.  A complementary analysis should also investigate whether a microscopic GUP realization modifies the interaction potential or quasiparticle spectrum, and should compare the resulting predictions with those of the effective phase-space prescription adopted here. Finally, a joint analysis to experimentally accessible thermodynamic observables, including their uncertainties and correlated systematic errors, would provide a more robust basis for constraining the deformation parameters.

In conclusion, Bose fluids with QQIs offer a theoretically interesting setting in which  to investigate how GUP-inspired modifications of phase space may influence finite-temperature many-body physics.
The present results establish analytical relationships between the deformation parameters and several thermodynamic observables within a clearly specified perturbative  scheme. 
They therefore provide a promising platform  for future quantitative  sensitivity studies, while demonstrating that reliable constraints on QG phenomenology require a consistent microscopic model, a realistic system, and justified uncertainties.




\section*{Research funding}
The authors received no financial support for the research, authorship, and publication of this article.

\newpage

\appendix
\include{AppendixA} 
 
\renewcommand{\theequation}{A\arabic{equation}}
\renewcommand{\thefigure}{A\arabic{figure}}

\section{ Angular integrals} \label{A}

In this Appendix we provide explicit formulas for angular integrals that contribute to zeroth-, first-, second-order GUP corrections to thermal depletion, critical temperature, 
and the EoS obtained in the main text.
We begin with integrals of the from $f_j(\epsilon_{\text{qq}})=\int_{0}^{\pi} \left[A^{-1}(\theta )+A(\theta )\right] A^{-j/2} (\theta) \sin\theta d \theta$.
\begin{align}
	f_3(\epsilon_{\text{qq}})&=\int_{0}^{\pi} \left(\frac{A (\theta)+A^{-1}(\theta)}{A^{3/2}(\theta)}\right) \sin\theta d \theta
	= \frac{2}{(1+8\epsilon_{\text{qq}})^{1/4}\left(-7+3\epsilon_q+72 \epsilon_{\text{qq}}^2\right)} 
\Bigg\{ -7-36\epsilon_q 
	+ \left(\frac{1+11\epsilon_{\text{qq}}+ 24 \epsilon_{\text{qq}}^2 }{(1+3\epsilon_{\text{qq}})^2}\right)^{1/4}  \\
&\times  \Bigg[ \left(-7+6\epsilon_{\text{qq}}(-7+12\epsilon_{\text{qq}})\right) 
F_1\!\left(\tfrac12;\tfrac14,\tfrac14;\tfrac32; a_-(\epsilon_{\text{qq}}), a_+ (\epsilon_{\text{qq}})\right) 
+ 70\,\epsilon_{\text{qq}}\, F_1\!\left(\tfrac32;\tfrac14,\tfrac14;\tfrac52;\,a_-(\epsilon_{\text{qq}}),\,a_+(\epsilon_{\text{qq}})\right) \Bigg] \Bigg\},\nonumber
\end{align}

\begin{align}
	f_5(\epsilon_{\text{qq}})&=\int_{0}^{\pi} \left(\frac{A (\theta)+A^{-1}(\theta)}{A^{5/2}(\theta)}\right) \sin\theta d \theta=\frac{2}{3(1+8\epsilon_{\text{qq}})^{1/4}\left(-7+3\epsilon_{\text{qq}}+72\epsilon_{\text{qq}}^2\right)} \Bigg\{7+36\epsilon_{\text{qq}}+(35-12\epsilon_{\text{qq}}(1+18\epsilon_{\text{qq}})) \nonumber\\
	&  \times b_+^{1/4} b_-^{3/4}  F_1\!\left(\tfrac12;\tfrac34,\tfrac32;\frac{2\sqrt{5\epsilon_{\text{qq}} (-7+24\epsilon_{\text{qq}})}} { b_+(\epsilon_{\text{qq}})
(1+3\epsilon_{\text{qq}}) }\right) \Bigg\},
\end{align}
and 
\begin{align}
	f_7(\epsilon_{\text{qq}})&=\int_{0}^{\pi} \left(\frac{A (\theta)+A^{-1}(\theta)}{A^{7/2}(\theta)}\right) \sin\theta d \theta
	=\frac{2}{(8 \epsilon_{\text{qq}} +1)^{5/4} \left(72 \epsilon_{\text{qq}} ^2+3 \epsilon_{\text{qq}}-7\right)^2} \\
&\times\bigg\{98+3 \epsilon_{\text{qq}} (427-12 \epsilon_{\text{qq}}
   (192 \epsilon_{\text{qq}} (3 \epsilon_{\text{qq}}+1)-101)) -\epsilon_{\text{qq}} (8\epsilon_{\text{qq}}+1) (24 \epsilon_{\text{qq}} (15\epsilon_{\text{qq}}+1)-77) \nonumber\\
&\times \left(\frac{1+11\epsilon_{\text{qq}}+ 24 \epsilon_{\text{qq}}^2 }{(1+3\epsilon_{\text{qq}})^2}\right)^{1/4}  \bigg[ 9 F_1\!\left(\tfrac12;\tfrac14,\tfrac14;\tfrac32; a_-(\epsilon_{\text{qq}}), a_+ (\epsilon_{\text{qq}})\right) 
-14 F_1\!\left(\tfrac12;\tfrac14,\tfrac14;\tfrac32; a_-(\epsilon_{\text{qq}}), a_+ (\epsilon_{\text{qq}})\right) \bigg] \bigg\} \nonumber.
\end{align}
where 
\[
a_{\pm}(\epsilon_{\text{qq}})= \frac{35\epsilon_{\text{qq}}}{15\epsilon_{\text{qq}} \pm\sqrt{5\epsilon_{\text{qq}}(-7+24\epsilon_{\text{qq}})}}, \quad \text{and} \quad
b_{\pm}(\epsilon_{\text{qq}})= \frac{1-12\epsilon_{\text{qq}}\pm \sqrt{5\epsilon_{\text{qq}} (-7+24\epsilon_{\text{qq}})} } {1+3\epsilon_{\text{qq}}}.
\]
\medskip
\noindent

For  integrals type $hj(\epsilon _{\text{qq}})=\int_{0}^{\pi}A^{-j/2} (\theta) \sin\theta d \theta$ that control the thermal contribution of the EoS and other thermodynamic quantities, 
one has:
\begin{align}
h_3(\epsilon_{\text{qq}}) =\int_{0}^{\pi} A^{-3/2} (\theta)\sin\theta d \theta
	= \frac{ 2}{(1+8\epsilon_{\text{qq}})^{3/4}}\left(1- a_{-}(\epsilon_{\text{qq}}) \right)^{3/4} \left(1+ a_{-}(\epsilon_{\text{qq}}) \right)^{1/4}
 F_1\!\left(\tfrac12;\tfrac34,\tfrac32;\frac{2\sqrt{5}\sqrt{\epsilon_q(-7+24\epsilon_q)}} {b_{+}(\epsilon_{\text{qq}})  (1+1+3\epsilon_{\text{qq}}) }\right),
\end{align}

\begin{align}
h_4(\epsilon_{\text{qq}})= \int_0^\pi A^{-2} (\theta)\sin\theta\,d\theta
= \sqrt{ \frac{7} {-7+24\epsilon_{\text{qq}}} }    \left[ \left( \frac{-a_+(\epsilon_{\text{qq}}) \arctan[-a_{+}(\epsilon_{\text{qq}})]}  {35\epsilon_{\text{qq}} }\right)  
- \left( \frac{-a_-(\epsilon_{\text{qq}}) \arctan[-a_{-}(\epsilon_{\text{qq}})]}  {35\epsilon_{\text{qq}} }\right)   \right]
\end{align}

\begin{align}
h_5(\epsilon_{\text{qq}})&=\int_{0}^{\pi} A^{-5/2}(\theta)  \sin\theta d \theta=-\frac{2}{(1 + 8 \epsilon_{\text{qq}})^{1/4} (-7 + 3 \epsilon_{\text{qq}} + 72 \epsilon_{\text{qq}}^2)}  
\Bigg [ 7 + 36 \epsilon_{\text{qq}}+ 45 \epsilon_{\text{qq}}  \left( \frac {35\epsilon_{\text{qq}}  (1+8\epsilon_{\text{qq}}) }   
{ 225 \epsilon_{\text{qq}}^2- 5 \epsilon_{\text{qq}}(24 \epsilon_{\text{qq}}-7) } \right)^{1/4} \\
       &\times F_1\left( \frac{1}{2}, \frac{1}{4}, \frac{1}{4}, \frac{3}{2}, a_{-}(\epsilon_{\text{qq}}) , a_{+}(\epsilon_{\text{qq}}) \right) 
 - 70 \epsilon_{\text{qq}}  \left( \frac {35\epsilon_q  (1+8\epsilon_q) }   { 225 \epsilon_q^2- 5 \epsilon_q(24 \epsilon_q-7) } \right)^{1/4}
 F_1\left( \frac{3}{2}, \frac{1}{4}, \frac{1}{4}, \frac{5}{2}, a_{-}(\epsilon_{\text{qq}}) , a_{+}(\epsilon_{\text{qq}})  \right) \Bigg] \nonumber,
\end{align}
and 
\begin{align}
h_7 (\epsilon_{\text{qq}})&=\int_{0}^{\pi} A^{-7/2} (\theta) \sin\theta d \theta=-\frac{2}{3(1 + 8 \epsilon_{\text{qq}})^{3/4} (-7 + 3\epsilon_{\text{qq}} + 72 \epsilon_{\text{qq}}^2)} \\
	&\times \Bigg[ -7 - 36 \epsilon_{\text{qq}} - 14 
	\left( \frac{-20 \epsilon_{\text{qq}}+ \sqrt{5} \sqrt{\epsilon_{\text{qq}} (-7 + 24 \epsilon_{\text{qq}})}}{15 \epsilon_{\text{qq}}
+ \sqrt{5} \sqrt{\epsilon_{\text{qq}} (-7 + 24 \epsilon_{\text{qq}})}} \right)^{1/4}
	\left( \frac{20 \epsilon_{\text{qq}} + \sqrt{5} \sqrt{\epsilon_{\text{qq}} (-7 + 24 \epsilon_{\text{qq}})}}{-15 \epsilon_{\text{qq}} + \sqrt{5} \sqrt{\epsilon_{\text{qq}} 
(-7 + 24 \epsilon_{\text{qq}}}} \right)^{3/4}  \nonumber\\
	&\quad \times F_1\left( \frac{1}{2}, \frac{3}{4}, \frac{3}{2},  a_{-}(\epsilon_{\text{qq}}), 
	\frac{2\sqrt{5}\sqrt{\epsilon_{\text{qq}}(-7+24\epsilon_{\text{qq}})}}  {b_{+}(\epsilon_{\text{qq}})  (1+1+3\epsilon_{\text{qq}}) } \right) 
        + 3 \epsilon_{\text{qq}} \left( \frac{-20 \epsilon_{\text{qq}} + \sqrt{5} \sqrt{\epsilon_{\text{qq}} (-7 + 24 \epsilon_{\text{qq}})}}{15 \epsilon_{\text{qq}} + \sqrt{5} \sqrt{\epsilon_{\text{qq}} (-7 + 24 \epsilon_{\text{qq}})}} \right)^{1/4} \nonumber\\
	&\times \left( \frac{20 \epsilon_{\text{qq}} + \sqrt{5} \sqrt{\epsilon_{\text{qq}} (-7 + 24 \epsilon_{\text{qq}})}}{-15\epsilon_{\text{qq}} + \sqrt{5} \sqrt{\epsilon_{\text{qq}} (-7 + 24 \epsilon_{\text{qq}})}} \right)^{3/4} 
           F_1\left( \frac{1}{2}, \frac{3}{4}, \frac{3}{2},\frac{2\sqrt{5} \sqrt{\epsilon_{\text{qq}} (-7 + 24 \epsilon_{\text{qq}})} }  {b_{+}(\epsilon_{\text{qq}})  (1+1+3\epsilon_{\text{qq}}) } \right) \Bigg]. \nonumber
\end{align}

 \begin{figure}
\centering 
\includegraphics[scale=0.75, angle=0] {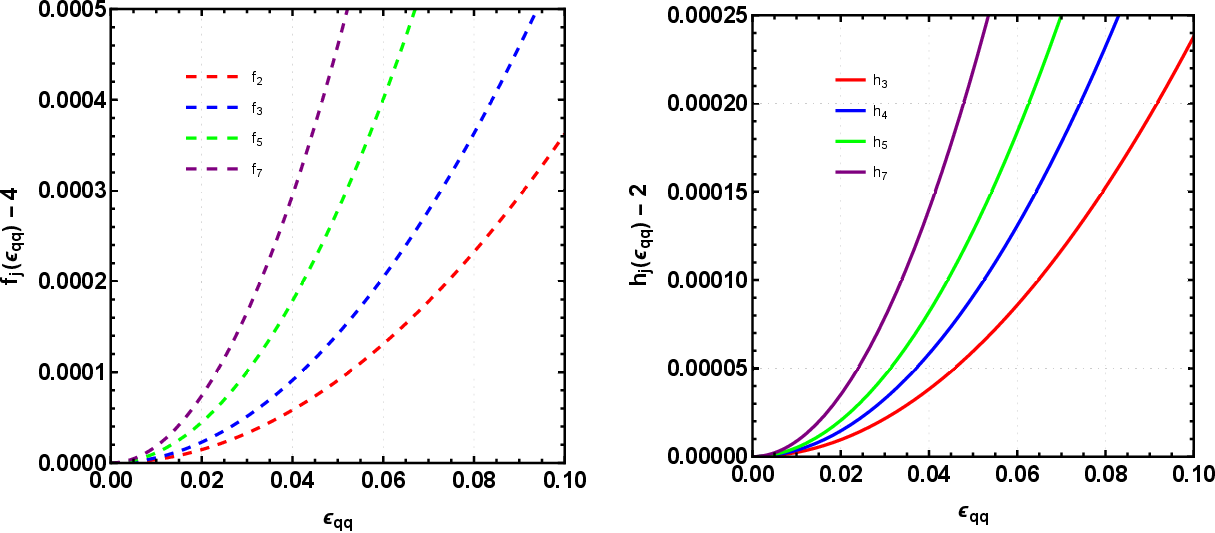} 
\caption { Angular functions  $f_j (\epsilon_{\text{qq}})$ and $h_j (\epsilon_{\text{qq}})$  as a function of  the quadrupolar coupling $\epsilon_{\text{qq}}$.
In the zero-coupling limit, one has  $f_j (0)=4$ and $h_j (0)=2$.}
\label{F5}
\end{figure}

The behavior of the  anisotropic functions $f_j (\epsilon_{\text{qq}})$ and $h_j (\epsilon_{\text{qq}})$ is displayed in Fig.~\ref{F5}. We see that they increase monotonically with $\epsilon_{\text{qq}}$ 
above their isotropic values ($f_j (0)=4$ and $h_j (0)=2$).
Functions with higher indices $j$ (e.g. $f_7$ and $h_7$) grow faster and experience significantly stronger amplification as $\epsilon_{\text{qq}}$  increases, reflecting the enhanced weigh placed on high-momentum anisotropic angular modes in higher-order GUP corrections.   

\include{appendixB} 
 
\renewcommand{\theequation}{B\arabic{equation}}
\renewcommand{\thefigure}{B\arabic{figure}}

\section{Other thermodynamic quantities} \label{B}


In this Appendix we extend our results obtained  within both GUP models to other relevant thermodynamic quantities. 
Starting from GUP equations   (\ref{QGUP2}), (\ref{DoS}) and  (\ref{DoSLQ}) and  the high-momentum Bogoliubov excitation energy (\ref{BogspecHT}) derived in the main text, 
we obtain analytical expressions for thermal contribution to the internal energy,  the heat capacity, and the entropy. 
The natural dimensionless thermodynamic quantities are then  $U/(nT_c)$, $C/n$, and $S/n$, where  $n$ is the total density.

  \subsection{Internal energy} 

In the grand canonical ensemble, the grand potential, $\Omega$,  can be written as: $\Omega= U-TS-\mu N$.
For thermal Bogoliubov excitations, where the effective chemical potential of the excitation gas is zero,  $\mu=0$, the internal energy  reduces directly to: 
  \begin{equation}  \label{InEgy}
  U=-\frac{\partial}{\partial \beta_T}\ln Z, \qquad \beta_T=\frac{1}{T},
   \end{equation} 
  Within the same  high-momentum Bogoliubov excitation energy (\ref{BogspecHT}) used in the main text, the internal energy for the QGUP case can be written as:
 \begin{equation}
{ \mathcal U}^{\rm QGUP}= \frac{U^{\rm QGUP}} {nT_c^{(0)}} = \frac{3\,\zeta(5/2)h_3(\epsilon_{\rm qq})} {4\,\zeta(3/2)f_3(\epsilon_{\rm qq})} \left(\frac{T}{T_c^{(0)}}\right)^{5/2} + \Delta U^{\rm QGUP}, 
 \end{equation} 
where
 \begin{equation}
\Delta \mathcal{U}^{\text{QGUP}} =  - \frac{45}{4} (\beta m T_c^{(0)}) \frac{\zeta(7/2)}{\zeta(3/2)} \frac{h_5(\epsilon_{\rm qq})}{f_3(\epsilon_{\rm qq})} \left(\frac{T} {T_c^{0} }\right)^{7/2} + \frac{315}{2} (\beta m T_c^{(0)})^2 
\frac{\zeta(9/2)}{\zeta(3/2)} \frac{h_7(\epsilon_{\rm qq})}{f_3(\epsilon_{\rm qq})} \left(\frac{T}{T_c^{(0)}}\right)^{9/2} .
\end{equation}
 
 Similarly, inserting LQGUP internal energy into Eq.~(\ref{InEgy}), gives
  \begin{equation}
{ \mathcal U}^{\rm LQGUP}= \frac{U^{\rm LQGUP}} {nT_c^{(0)}} = \frac{3\,\zeta(5/2)h_3(\epsilon_{\rm qq})} {4\,\zeta(3/2)f_3(\epsilon_{\rm qq})} \left(\frac{T}{T_c^{(0)}}\right)^{5/2} + \Delta U^{\rm LQGUP}, 
 \end{equation} 
where
\begin{equation}
\Delta \mathcal{U}^{\text{LQGUP}}  =  \alpha \sqrt{m T_c^{(0)}} \frac{32 \,\zeta(3)}{ \sqrt{\pi} \zeta(3/2)} \frac{h_4(\epsilon_{\rm qq})}{f_3(\epsilon_{\rm qq})} \left(\frac{T}{T_c^{(0)}}\right)^3 - \beta m T_c^{(0)} \frac{375\,\zeta(7/2)}{8\,\zeta(3/2)}
 \frac{h_5(\epsilon_{\rm qq})}{f_3(\epsilon_{\rm qq})} \left(\frac{T}{T_c^{(0)}}\right)^{7/2} .
\end{equation}

  \subsection{Heat capacity} 
  
 The heat capacity at constant volume is defined by 
 \begin{equation} C= \left(\frac{\partial U}{\partial T}\right)_V. 
 \end{equation} 
 Differentiation with respect to temperature gives
\begin{equation} 
\frac{C}{n} = \frac{d}{d (T/T_c)} \left( \frac{U}{nT_c} \right). 
\end{equation} 
This normalization is dimensionless and follows directly from the definition of the heat capacity.

For QGUP, differentiation of Eq. (B1)  gives 
\begin{equation} 
{\mathcal C}^{\rm QGUP}= \frac{C^{\rm QGUP}}{n} = \frac{15\zeta(5/2)h_3(\epsilon_{\rm qq})} {8\zeta(3/2)f_3 (\epsilon_{\rm qq})}  \left(\frac{T}{T_c^{(0)}}\right)^{3/2} + \Delta C_{\rm QGUP}, 
 \end{equation} 
where
 \begin{equation}
\Delta \mathcal{C}^{\text{QGUP}} = - \frac{315}{8} (\beta m T_c^{(0)}) \frac{\zeta(7/2)}{\zeta(3/2)} \frac{h_5(\epsilon_{\rm qq})}{f_3(\epsilon_{qq})} \left(\frac{T}{T_c^{(0)}}\right)^{5/2} 
+ \frac{2835}{4} (\beta m T_c^{(0)})^2 \frac{\zeta(9/2)}{\zeta(3/2)} \frac{h_7(\epsilon_{\rm qq})}{f_3(\epsilon_{\rm qq})} \left(\frac{T}{T_c^{(0)}}\right)^{7/2}.
\end{equation}
 
 For LQGUP, the corresponding heat capacity is 
\begin{equation}
{\mathcal C}^{\rm LQGUP}= \frac{C^{\rm LQGUP}}{n} = \frac{15\,\zeta(5/2)h_3(\epsilon_{\rm qq})} {8\,\zeta(3/2)f_3(\epsilon_{\rm qq})}  \left(\frac{T}{T_c^{(0)}}\right)^{3/2} + \Delta C^{\rm LQGUP}, 
\end{equation} 
with
\begin{equation}
\Delta \mathcal{C}^{\text{LQGUP}}= \alpha \sqrt{m T_c^{(0)}} \frac{96 \zeta(3)}{\sqrt{\pi} \zeta(3/2)} \frac{h_4(\epsilon_{\rm qq})}{f_3(\epsilon_{\rm qq})} \left(\frac{T}{T_c^{(0)}}\right)^2 - \beta m T_c^{(0)} \frac{2625\,\zeta(7/2)}{16\,\zeta(3/2)} \frac{h_5(\epsilon_{\rm qq})}{f_3(\epsilon_{\rm qq})} \left(\frac{T}{T_c^{(0)}}\right)^{5/2}. 
\end{equation}

 \subsection{Entropy} 
 Once the GUP-corrected heat capacity has been obtained, the corresponding entropy follows directly by integrating $\mathcal C/n$ with respect to temperature, 
  \begin{equation} 
  S(T) = \int_0^T \frac{C(T')}{T'}\,dT', 
  \end{equation} 
  For QGUP, the entropy takes the form
  \begin{equation}
  {\mathcal S}^{\rm QGUP}= \frac{S^{\rm QGUP}}{n} = \frac{5\,\zeta(5/2)h_3(\epsilon_{\rm qq})} {4\,\zeta(3/2)f_3(\epsilon_{\rm qq})} \left(\frac{T}{T_c^{(0)}}\right)^{3/2} + \Delta S^{\rm QGUP}, 
  \end{equation} 
 where 
 \begin{equation}
\Delta \mathcal{S}^{\text{QGUP}}= - \frac{ 63}{4} \,(\beta m T_c^{(0)}) \frac{\zeta(7/2)}{\zeta(3/2)} \frac{h_5(\epsilon_{\rm qq})}{f_3(\epsilon_{\rm qq})} \left(\frac{T}{T_c^{(0)}}\right)^{5/2} 
+\, \frac{405}{4} (\beta m T_c^{(0)})^2 \frac{\zeta(9/2)}{\zeta(3/2)} \frac{h_7(\epsilon_{\rm qq})}{f_3(\epsilon_{\rm qq})} \left(\frac{T}{T_c^{(0)}}\right)^{7/2}. 
\end{equation}

 For the LQGUP, one obtains
  \begin{equation} \frac{S_{\rm LQGUP}}{n} = \frac{5\,\zeta(5/2)h_3(\epsilon_{\rm qq})} {4\,\zeta(3/2)f_3(\epsilon_{\rm qq})} \left(\frac{T}{T_c^{(0)}}\right)^{3/2} + \Delta S_{\rm LQGUP}, 
   \end{equation} 
 where 
 \begin{equation}
\Delta \mathcal{S}^{\text{LQGUP}}=  \alpha \sqrt{m T_c^{(0)}} \frac{48\,\zeta(3)}{\sqrt{\pi}\zeta(3/2)} \frac{h_4(\epsilon_{\rm qq})}{f_3(\epsilon_{\rm qq})} \left(\frac{T}{T_c^{(0)}}\right)^2 - \beta m T_c^{(0)} \frac{525\,\zeta(7/2)}{8\zeta(3/2)} 
\frac{h_5(\epsilon_{\rm qq})}{f_3(\epsilon_{\rm qq})} \left(\frac{T}{T_c^{(0)}}\right)^{5/2}. 
\end{equation}
 Remarkably, the leading temperature dependence are: $ { \mathcal U} \propto \left(T/T_c\right)^{5/2}$,  ${ \mathcal C}   \propto \left(T/T_c\right){3/2}$, and  ${ \mathcal S}  \propto \left(T/T_c\right)^{3/2}$, as required by thermodynamics consistency.

 \begin{figure}
\centering 
\includegraphics[scale=0.7, angle=0] {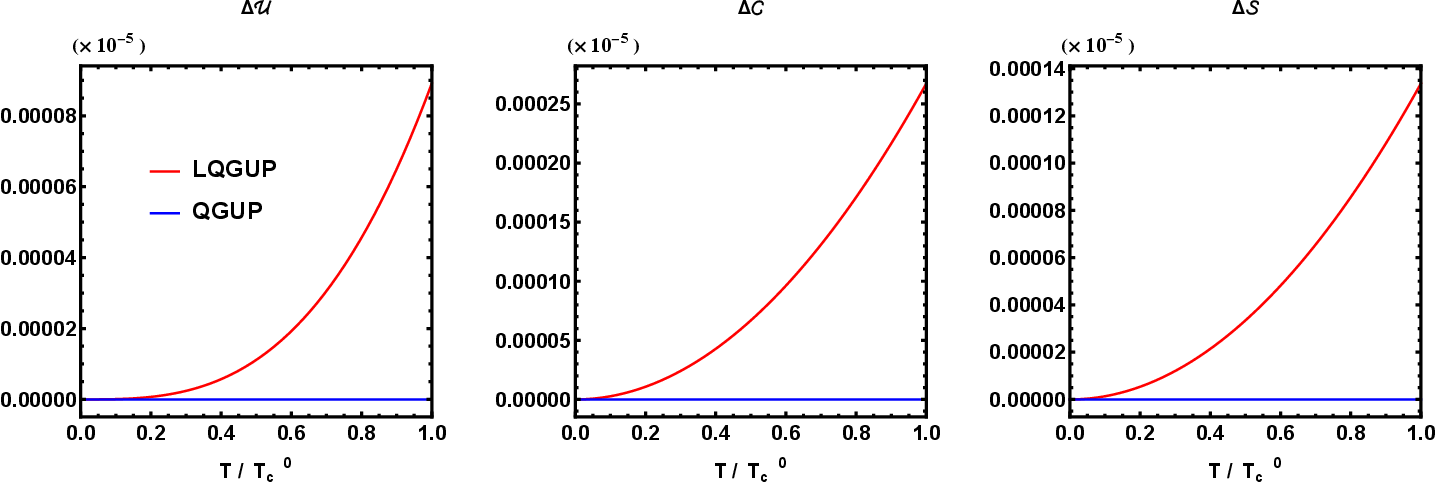}
\caption {Comparison between  QGUP and LQGUP contributions to (a) the thermal-internal-energy correction $\Delta \mathcal{U}$, (b) the heat-capacity correction $\Delta \mathcal{C}$, (c) the entropy correction $\Delta \mathcal{S}$,
as a function of the  critical temperature,  $T /Tc^{(0)}$.}
\label{F6}
\end{figure}

 \subsection{Discussion} 
 The corrected expressions for the internal energy,  heat capacity, and  entropy  are mutually consistent order by order  in the QGUP and LQGUP parameters. 
 Their relative magnitudes and temperature dependences rely on the QQI through $f_j(\epsilon_{\rm qq})$ and $h_j(\epsilon_{\rm qq})$, as well as  on the dimensionless combinations   $\beta mT_c$ and $\alpha(mT_c)^{1/2}$. 
 Therefore, no general ordering of the QGUP and LQGUP corrections is imposed independently of the parameters. For the same reason, the numerical results shown in Fig.~\ref{F6} should be interpreted as a comparison of the magnitude and temperature  dependence of  the corrections for the particular parameter values used in the calculation, rather than as a universal statement that one GUP model always produces larger thermodynamic modifications than the other.
 The corrected expressions also preserve the expected low-temperature behavior within the validity range of the approximation.
 In particular, as   $T\rightarrow0$,  ${ \mathcal U} $,  ${ \mathcal C}$, and ${ \mathcal S} $ vanish. These result should nevertheless be interpreted within the high-momentum Bogoliubov regime and within the perturbative domain in which the GUP expansion is controlled.

\end{document}